\documentclass[aps,prd,preprint,superscriptaddress,tightenlines,nofootinbib]{revtex4}

\usepackage{graphicx}
\usepackage{dcolumn}
\usepackage{bm}

\def\kpi{K^-\pi^+}
\def\kpipiz{K^-\pi^+\pi^0}
\def\kpipipi{K^-\pi^+\pi^+\pi^-}
\def\kpipi{K^-\pi^+\pi^+}
\def\kpipipiz{K^-\pi^+\pi^+\pi^0}
\def\kspi{K^0_S\pi^+}
\def\kspipiz{K^0_S\pi^+\pi^0}
\def\kspipipi{K^0_S\pi^+\pi^+\pi^-}
\def\kkpi{K^+K^-\pi^+}

\def\Dzkpi{D^0\to K^-\pi^+}
\def\Dzkpipiz{D^0\to K^-\pi^+\pi^0}
\def\Dzkpipipi{D^0\to K^-\pi^+\pi^+\pi^-}
\def\Dpkpipi{D^+\to K^-\pi^+\pi^+}
\def\Dpkpipipiz{D^+\to K^-\pi^+\pi^+\pi^0}
\def\Dpkspi{D^+\to K^0_S\pi^+}
\def\Dpkspipiz{D^+\to K^0_S\pi^+\pi^0}
\def\Dpkspipipi{D^+\to K^0_S\pi^+\pi^+\pi^-}
\def\Dpkkpi{D^+\to K^+K^-\pi^+}

\def\Dzbarkpi{\bar D^0\to K^+\pi^-}
\def\Dzbarkpipiz{\bar D^0\to K^+\pi^-\pi^0}
\def\Dzbarkpipipi{\bar D^0\to K^+\pi^-\pi^-\pi^+}
\def\Dmkpipi{D^-\to K^+\pi^-\pi^-}
\def\Dmkpipipiz{D^-\to K^+\pi^-\pi^-\pi^0}
\def\Dmkspi{D^-\to K^0_S\pi^-}
\def\Dmkspipiz{D^-\to K^0_S\pi^-\pi^0}
\def\Dmkspipipi{D^-\to K^0_S\pi^-\pi^-\pi^+}
\def\Dmkkpi{D^-\to K^+K^-\pi^-}

\def\NDDbar{N_{D\bar D}}
\def\NDzDzbar{N_{D^0\bar D^0}}
\def\NDpDm{N_{D^+D^-}}

\begin{document}

\preprint{CLNS 05/1914}       
\preprint{CLEO 05-6}         

\title{\boldmath Measurement of Absolute Hadronic Branching Fractions of $D$
Mesons and $e^+e^-\to D\bar D$ Cross Sections at $E_{\rm cm}=3773$ MeV}

\author{Q.~He}
\author{H.~Muramatsu}
\author{C.~S.~Park}
\author{W.~Park}
\author{E.~H.~Thorndike}
\affiliation{University of Rochester, Rochester, New York 14627}
\author{T.~E.~Coan}
\author{Y.~S.~Gao}
\author{F.~Liu}
\affiliation{Southern Methodist University, Dallas, Texas 75275}
\author{M.~Artuso}
\author{C.~Boulahouache}
\author{S.~Blusk}
\author{J.~Butt}
\author{E.~Dambasuren}
\author{O.~Dorjkhaidav}
\author{J.~Li}
\author{N.~Menaa}
\author{R.~Mountain}
\author{R.~Nandakumar}
\author{K.~Randrianarivony}
\author{R.~Redjimi}
\author{R.~Sia}
\author{T.~Skwarnicki}
\author{S.~Stone}
\author{J.~C.~Wang}
\author{K.~Zhang}
\affiliation{Syracuse University, Syracuse, New York 13244}
% ...
\collaboration{CLEO Collaboration} 
\noaffiliation

\date{April 1, 2005}

\begin{abstract} 
Using 55.8 pb$^{-1}$ of $e^+e^-$ collisions recorded at the $\psi(3770)$
resonance with the CLEO-c detector at CESR, we determine
absolute hadronic branching fractions of charged and neutral
$D$ mesons using a double tag technique. Among measurements for three
$D^0$ and six $D^+$ modes, we obtain reference branching fractions
${\cal B}(D^0\to K^-\pi^+)=(3.91\pm 0.08\pm 0.09)\%$ and
${\cal B}(D^+\to K^-\pi^+\pi^+)=(9.5\pm 0.2\pm 0.3)\%$, where the
uncertainties are statistical and systematic, respectively.  
Final state radiation is included in these branching
fractions by allowing for additional, unobserved, photons in the 
final state.
Using a
determination of the integrated luminosity, we also extract the cross sections 
$\sigma(e^+e^-\to D^0\bar D^0)=(3.60\pm 0.07^{+0.07}_{-0.05})\ {\rm nb}$ and 
$\sigma(e^+e^-\to D^+D^-)=(2.79\pm 0.07^{+0.10}_{-0.04})\ {\rm nb}$. 
\end{abstract}

\maketitle

Absolute measurements of hadronic charm meson branching
fractions play a central role in the study of the weak
interaction because they serve to normalize many $D$ and $B$
meson branching fractions, from which elements of the
Cabibbo-Kobayashi-Maskawa (CKM) matrix~\cite{ckm} are determined.
For instance, the determination of the CKM matrix element $|V_{cb}|$ from the
$B\to D^*\ell\nu$ decay rate using full $D^*$ reconstruction requires 
knowledge of the
$D$ meson branching fractions~\cite{vcbreview}.  In this Letter,
we present charge-averaged branching fraction measurements of
three $D^0$ and six $D^+$ decay modes: $\Dzkpi$, $\Dzkpipiz$, $\Dzkpipipi$,
$\Dpkpipi$, $\Dpkpipipiz$, $\Dpkspi$, $\Dpkspipiz$, $\Dpkspipipi$,
and $\Dpkkpi$.
Two of these modes, $\Dzkpi$ and $\Dpkpipi$,
are particularly important because essentially all other
$D^0$ and $D^+$ branching fractions have been determined from
ratios to one of these branching fractions~\cite{PDG}.

To date, the most precise measurements of hadronic $D$ branching fractions
are made with slow-daughter-pion tagging of $D^*$ mesons 
from $Z^0$ decays and from continuum production 
in $e^+e^-$ interactions at the $\Upsilon(4S)$~\cite{aleph,cleoII}.
Previously, the MARK III collaboration 
measured hadronic branching fractions at the $D\bar D$ threshold
using a double tagging technique which
relied on fully-reconstructed $\psi(3770)\to D\bar D$
decays~\cite{markiii-1, markiii-2}.
This technique obviated the need for knowledge of the luminosity or the
$e^+e^-\to D\bar D$ production cross section.
We employ a similar technique using CLEO-c data,
in a sample roughly six times larger than that of MARK III, resulting
in precision comparable to the current PDG world averages.

The data sample we analyze was produced in
$e^+e^-$ collisions at the Cornell Electron Storage Ring (CESR)
and collected with the CLEO-c detector.
It consists of 55.8 ${\rm pb}^{-1}$ of integrated luminosity
collected on the $\psi(3770)$ resonance, at
a center-of-mass energy $E_{\rm cm}=3773$ MeV.
At this energy, no additional hadrons accompanying the
$D\bar D$ pairs are produced.  Reconstruction of one $D$ or
$\bar D$ meson (called single tag or ST) tags the event
as either $D^0\bar D^0$ or $D^+D^-$.  For a given decay mode $i$, we
measure independently the $D$ and $\bar D$ ST yields,
denoted by $N_i$ and $\bar N_i$. We determine the corresponding efficiencies,
denoted by $\epsilon_i$ and $\bar\epsilon_i$,
from Monte Carlo simulations.
Thus, $N_i=\epsilon_i{\cal B}_iN_{D\bar D}$ and
$\bar N_i=\bar\epsilon_i{\cal B}_i N_{D\bar D}$, 
where ${\cal B}_i$ is the branching fraction for
mode $i$,  assuming no $CP$ violation, and $N_{D\bar D}$ is the 
number of produced $D\bar D$ pairs.  Double tag (DT) events are the subset of ST events
where both the $D$ and $\bar D$ are reconstructed.  The DT
yield for $D$ mode $i$ and $\bar D$ mode $j$, denoted by
$N_{ij}$, is given by
$N_{ij} = \epsilon_{ij}{\cal B}_i{\cal B}_jN_{D\bar D}$,
where $\epsilon_{ij}$ is the DT efficiency.  As with ST yields,
the charge conjugate DT yields and efficiencies, $N_{ji}$ and
$\epsilon_{ji}$, are determined separately.  Charge conjugate
particles are implied, unless referring to ST and DT yields.

 The ${\cal B}_i$ can be determined from the DT yield $N_{ij}$
and the corresponding ST yield $\bar N_j$ via
${\cal B}_i = [N_{ij}/\bar N_j] \times [\bar\epsilon_j/\epsilon_{ij}]$.
Similarly, we have $\NDDbar = [N_i\bar N_j/N_{ij}] \times [\epsilon_{ij}/(\epsilon_i\bar\epsilon_j)]$.
Because $\epsilon_{ij}\approx\epsilon_i\bar\epsilon_j$,
the branching fractions thus obtained are nearly independent
of the tag mode efficiency, and $\NDDbar$ is nearly independent of all
efficiencies.
We extract branching fractions and $\NDDbar$ by combining
ST and DT yields with a least squares technique.
Although the $D^0$ and $D^+$ yields are statistically independent,
systematic effects and misreconstruction resulting in crossfeed introduce
correlations among their uncertainties.  Therefore,
we fit $D^0$ and $D^+$ parameters simultaneously, including in the $\chi^2$
statistical and systematic uncertainties and their correlations for all
experimental inputs~\cite{brfit}.
Thus, yields, efficiencies, and backgrounds are treated uniformly, and
the statistical uncertainties on ${\cal B}_i$ and $\NDDbar$
include the correlations among $N_i$, $\bar N_j$, and $N_{ij}$.
Also, in the above efficiency ratios most systematic uncertainties
are correlated between ST and DT efficiencies,
so their effects largely cancel.

The CLEO-c detector is a modification of the CLEO~III
detector~\cite{cleoiidetector,cleoiiidr,cleorich}, in which the silicon-strip
vertex detector was replaced with a six-layer vertex drift chamber, whose
wires are all at small stereo angles to the beam axis~\cite{cleocyb}.  The
charged particle tracking system, consisting of this vertex drift chamber and
a 47-layer central drift chamber~\cite{cleoiiidr} operates in a 1.0~T
magnetic field, oriented along the beam axis. The momentum resolution
of the tracking system is approximately 0.6\% at
$p=1$~GeV/$c$. Photons are detected in an electromagnetic calorimeter,
composed of 7800 CsI(Tl) crystals~\cite{cleoiidetector}, which
attains a photon energy resolution of 2.2\% at $E_\gamma=1$~GeV
and 5\% at 100~MeV. The solid angle coverage for charged and neutral
particles of the CLEO-c detector is 93\% of $4\pi$.  We utilize two
particle identification (PID) devices to separate $K^\pm$ from $\pi^\pm$: 
the central drift chamber, which provides measurements of ionization energy
loss ($dE/dx$), and, surrounding this drift chamber, a cylindrical
ring-imaging Cherenkov (RICH) detector~\cite{cleorich}, whose active 
solid angle is 80\% of $4\pi$.  The combined
$dE/dx$-RICH PID system has a pion or kaon efficiency
$>90$\% and a probability of pions faking kaons (or vice versa) $<5$\%.
The response of the experimental apparatus is studied with a detailed
GEANT-based~\cite{geant} Monte Carlo simulation of the CLEO detector
for particle trajectories generated by EvtGen~\cite{evtgen}
and final state radiation (FSR) predicted by PHOTOS~\cite{photos}.
Simulated events are processed in a fashion similar to data.
The data sample's integrated luminosity ($\cal{L}$) is measured 
using $e^+e^-$ Bhabha events in the calorimeter~\cite{lumins}, where the event
count normalization is provided by the detector simulation.

Charged tracks are required to be well-measured and to satisfy 
criteria based on the track fit quality.  They must also be consistent with
coming from the interaction point in three dimensions. Pions and kaons are
identified by consistency with the expected $dE/dx$ and RICH information,
when available.  We form $\pi^0$ candidates from
photon pairs with invariant mass within 3 standard deviations ($\sigma$),
with $\sigma\approx 5$--7 MeV/$c^2$ depending on photon energy and location,
of the known $\pi^0$ mass.
These candidates are then fit kinematically with their masses
constrained to the known $\pi^0$ mass.
The $K^0_S$ candidates are selected from pairs of oppositely-charged and
vertex-constrained tracks having invariant mass within 12 MeV/$c^2$,
or roughly $4.5\sigma$, of the known $K^0_S$ mass.

We identify $D$ meson candidates by their invariant masses and total
energies.  We calculate a beam-constrained mass by
substituting the beam energy, $E_0$, for the measured $D$
candidate energy: $Mc^2 \equiv\sqrt{E_0^2 - {\mathbf p}_D^2c^2}$, where
${\mathbf p}_D$ is the $D$ candidate momentum.  Performing this substitution
improves the resolution of $M$ by one order of magnitude, to about
2 MeV/$c^2$, which is dominated by the beam
energy spread.  We define $\Delta E\equiv E_D - E_0$, where
$E_D$ is the sum of the $D$ candidate daughter energies.
For final states consisting entirely of tracks, the $\Delta E$
resolution is 7--10 MeV.  A $\pi^0$
in the final state degrades this resolution by roughly a factor of two.
We accept $D$ candidates with $M$ greater than 1.83 GeV/$c^2$ and with
mode-dependent $\Delta E$ requirements of approximately $3\sigma$.
For both ST and DT modes, we accept at most one candidate per mode
per event.  In ST modes, the candidate with the smallest $\Delta E$
is chosen, while in DT modes, we take the candidate whose average
of $D$ and $\bar D$ $M$ values, denoted by $\widehat{M}$,
is closest to the known $D$ mass.

We extract ST and DT yields from $M$ distributions in the
samples described above. We perform unbinned maximum likelihood fits
in one and two dimensions for ST and DT modes, respectively,
to a signal shape and one
or more background components. The signal shape includes the effects of 
beam energy smearing, initial state radiation, the
line shape of the $\psi(3770)$, and reconstruction resolution.
The background in ST modes is described by an
ARGUS function~\cite{argusf}, which models combinatorial
contributions.  In DT modes, backgrounds can be uncorrelated,
where either the $D$ or $\bar D$ is misreconstructed,
or correlated, where all the final state particles in the event are correctly
reconstructed but are mispartitioned among the $D$ and $\bar D$.
In fitting the two-dimensional $M(D)$ versus $M(\bar D)$
distribution, we model the uncorrelated background by a pair of
functions, where one dimension is an ARGUS function and the other
is the signal shape.  We model the correlated background 
by an ARGUS function in $\widehat{M}$ and a Gaussian in the
orthogonal variable, which is $[M(\bar D)-M(D)]/2$.

\begin{table}[tb]
\caption{Single tag data yields and efficiencies and their
statistical uncertainties.}
\label{tab-STYieldsAndEffs}
\begin{center}
\begin{tabular}{lcc}
\hline\hline
$D$ or $\bar D$ Mode &  Yield ($10^3$) & Efficiency (\%) \\ \hline
$\Dzkpi$               &  $5.11\pm 0.07$ & $64.6\pm 0.3$ \\
$\Dzbarkpi$            &  $5.15\pm 0.07$ & $65.6\pm 0.3$ \\
$\Dzkpipiz$            &  $9.51\pm 0.11$ & $31.4\pm 0.1$ \\
$\Dzbarkpipiz$         &  $9.47\pm 0.11$ & $31.8\pm 0.1$ \\
$\Dzkpipipi$           &  $7.44\pm 0.09$ & $43.6\pm 0.2$ \\
$\Dzbarkpipipi$        &  $7.43\pm 0.09$ & $43.9\pm 0.2$ \\ \hline
$\Dpkpipi$             &  $7.56\pm 0.09$ & $50.7\pm 0.2$ \\
$\Dmkpipi$             &  $7.56\pm 0.09$ & $51.3\pm 0.2$ \\
$\Dpkpipipiz$          &  $2.45\pm 0.07$ & $25.7\pm 0.2$ \\
$\Dmkpipipiz$          &  $2.39\pm 0.07$ & $25.7\pm 0.2$ \\
$\Dpkspi$              &  $1.10\pm 0.04$ & $45.5\pm 0.4$ \\
$\Dmkspi$              &  $1.13\pm 0.04$ & $45.9\pm 0.4$ \\
$\Dpkspipiz$           &  $2.59\pm 0.07$ & $22.4\pm 0.2$ \\
$\Dmkspipiz$           &  $2.50\pm 0.07$ & $22.4\pm 0.2$ \\
$\Dpkspipipi$          &  $1.63\pm 0.06$ & $31.1\pm 0.2$ \\
$\Dmkspipipi$          &  $1.58\pm 0.06$ & $31.3\pm 0.2$ \\
$\Dpkkpi$              &  $0.64\pm 0.03$ & $41.4\pm 0.5$ \\
$\Dmkkpi$              &  $0.61\pm 0.03$ & $40.8\pm 0.5$ \\
\hline\hline
\end{tabular}
\end{center}
\end{table}

\begin{figure}
\includegraphics*[width=\linewidth]{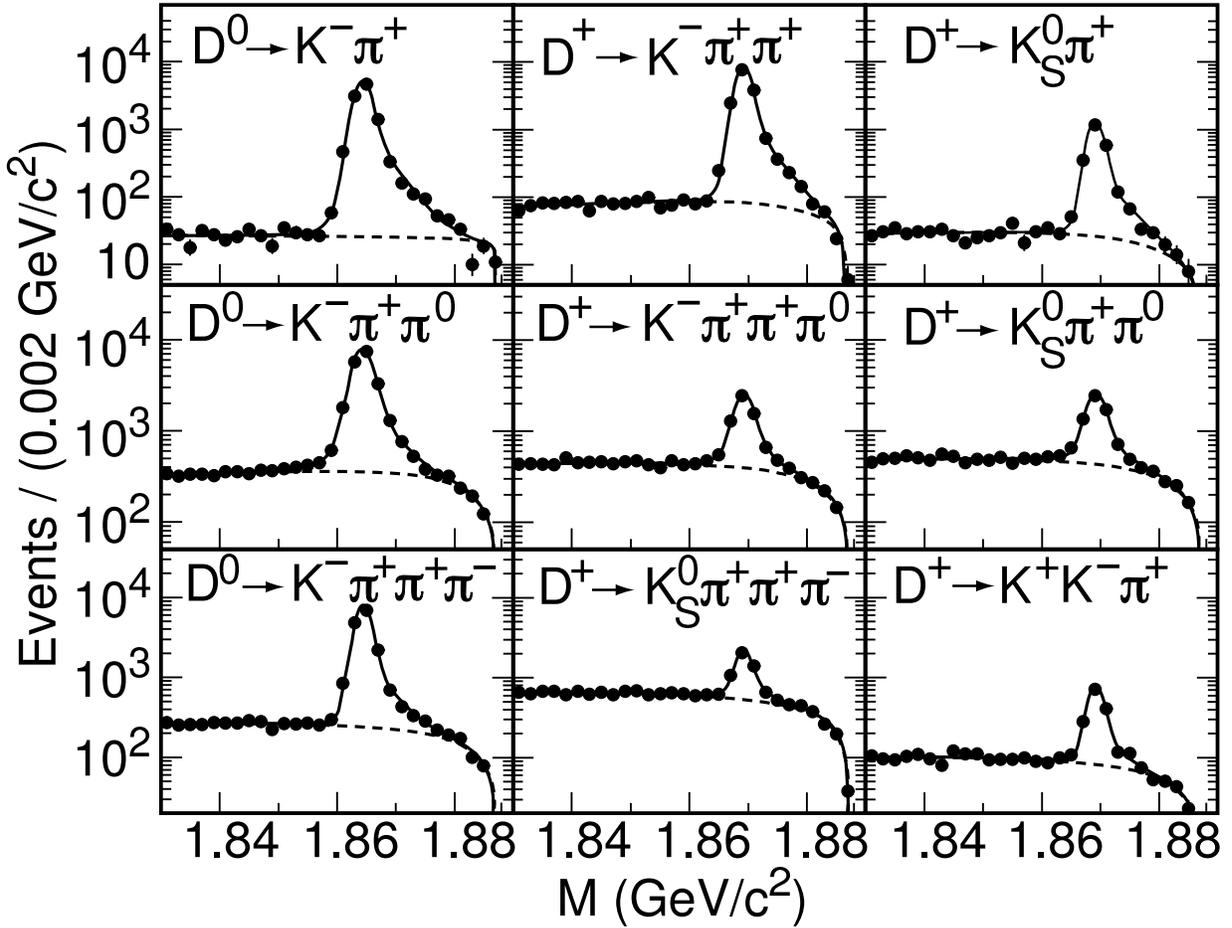}
\caption{
Semilogarithmic plots of ST yields and fits, with $D$ and $\bar D$ combined
for each mode.  Data are shown as points with error bars.  The solid lines show
the total fits and the dashed lines the background shapes. The high mass tails
on the signal are due to initial state radiation.
}
\label{fig-ST}
\end{figure}

\begin{figure}
\includegraphics*[width=0.99\linewidth]{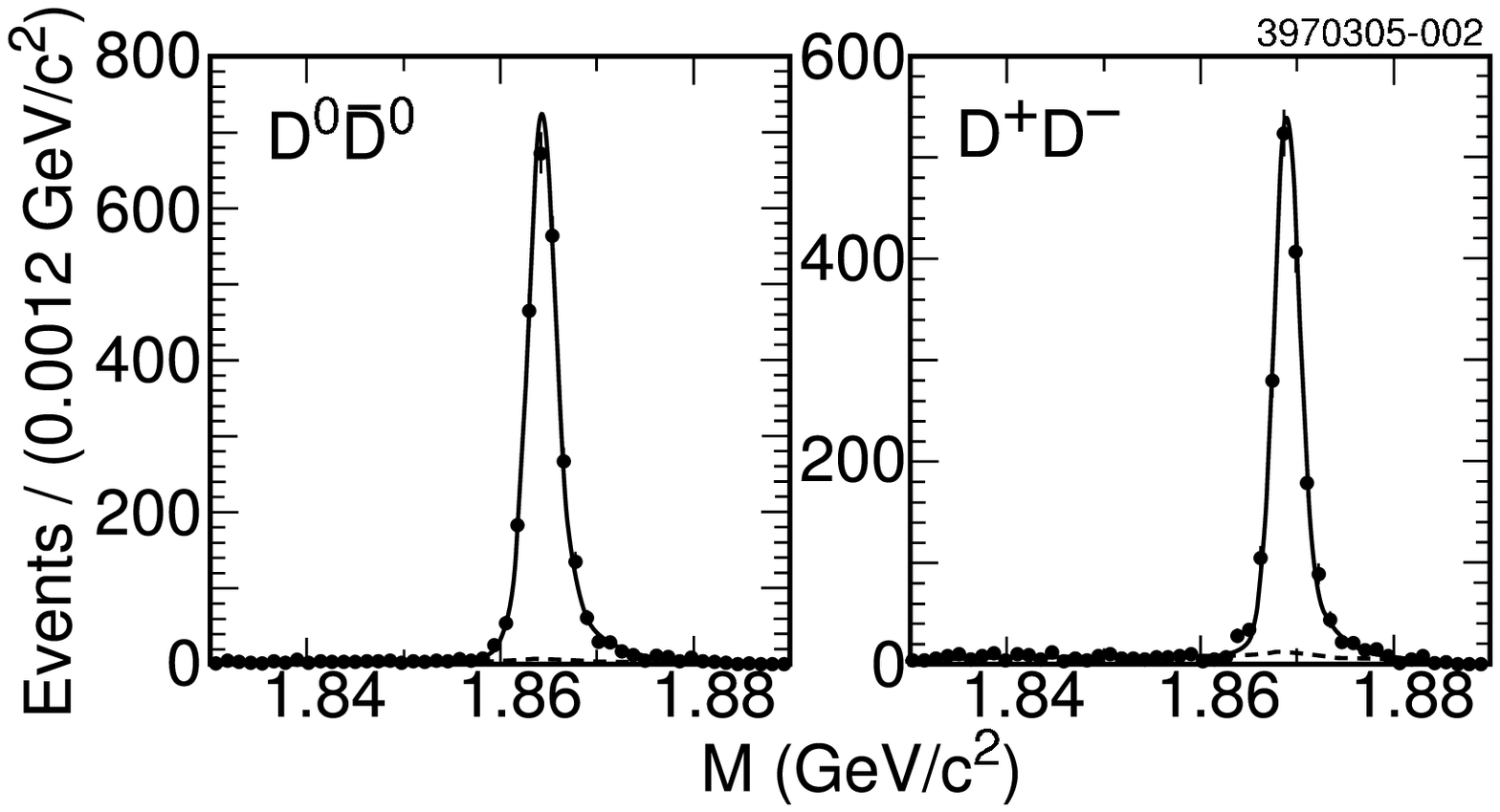}
\caption{
DT yields and fits projected onto the $D^0$ and $D^+$ axes and summed over all
modes.  Data are shown as points with error bars.  The solid lines show the
total fit and the dashed lines the background shapes.
}
\label{fig-DT}
\end{figure}

Table~\ref{tab-STYieldsAndEffs} gives the 18 ST data yields and efficiencies
determined from simulated events.  Figure~\ref{fig-ST} shows the $M$ distributions for the nine
decay modes with $D$ and $\bar D$ candidates combined.  Overlaid are the
fitted signal and background components.
We also measure 45 DT yields in data and determine the
corresponding efficiencies from simulated events.  Figure~\ref{fig-DT} shows
$M(D)$ for all modes combined, separated by
charge.  We find total DT yields of $2484\pm 51$ for $D^0$ and
$1650\pm 42$ for $D^+$.
Because of the cleanliness of the DT modes, their statistical yield
uncertainties are close to $\sqrt{N_{ij}}$.

Using a missing mass technique, we measure efficiencies for
reconstructing tracks, $K^0_S$ decays, and $\pi^0$ decays in both
data and simulated events.  We fully reconstruct $\psi(3770)\to D\bar D$,
$\psi(2S)\to J/\psi\pi^+\pi^-$, and $\psi(2S)\to J/\psi\pi^0\pi^0$
events, leaving out one particle, for which we wish to determine
the efficiency.   The missing mass of this combination peaks at
the mass of the omitted particle, whether or not it is detected.
Then, the desired efficiency is the fraction of this peak with this
particle correctly reconstructed.  For tracks and $K^0_S$ candidates, 
we find good
agreement between efficiencies in data and simulated events.  For
$\pi^0$ candidates, we correct the 
simulated efficiencies by 3.9\%, which is the level
of disagreement with data found in this study.  The
relative uncertainties in these determinations, 0.7\% per track,
3.0\% per $K^0_S$, and 2.0\% per $\pi^0$, are the largest contributions
to the systematic uncertainties.

We study the simulation of the PID efficiencies using decays with
unambiguous particle content, such as $D^0\to K^0_S\pi^+\pi^-$ and
$D^+\to K^-\pi^+\pi^+$.  We find a need to correct the simulated
efficiencies by 0.3\% for $\pi^\pm$ and 1.3\% for $K^\pm$, and we apply
associated systematic uncertainties of the same size.
Other sources of efficiency uncertainty include: 
the $\Delta E$ requirements (1.0--2.5\%), for which we examine
$\Delta E$ sidebands; modeling of particle multiplicity and detector 
noise (0.2--1.3\%);
and modeling of resonant substructure in multi-body
modes (0.4--1.5\%), which we assess by comparing
simulated momentum spectra to those in data.  We also include additive
uncertainties of 0.5\% to account for variations of yields with 
fit function.
Smaller systematic uncertainties arise from online and offline filtering
(0.4\%), simulation of FSR (0.5\% per $D$ or $\bar D$), and the assumed
width of the $\psi(3770)$ in the $M$ signal shape (0.6\%).
The effect of quantum correlations between the $D^0$ and $\bar D^0$ states
appears through $D^0$-$\bar D^0$ mixing and through doubly
Cabibbo-suppressed decays~\cite{CPCorrelation}.  The former contribution
is limited by available measurements~\cite{PDG} to be less than
${\cal O}(10^{-3})$ and is, therefore, neglected in this analysis.
The latter contribution is addressed with a systematic uncertainty due to
the unknown phase of interference in neutral DT modes between the
Cabibbo-favored amplitude and the amplitude for doubly Cabibbo-suppressed
transitions in both $D^0$ and $\bar D^0$ (0.8\%).

The branching fraction fitter~\cite{brfit} takes these systematic
uncertainties as input, along with ST and DT yields and efficiencies,
crossfeed probabilities among the modes, background branching fractions
and efficiencies, and statistical uncertainties on all of these measurements.
The estimated crossfeed and
background contributions produce yield adjustments of ${\cal O}(1\%)$.
Their dependence on the fit parameters is
taken into account both in the yield subtraction and in the $\chi^2$
minimization.
We validated the analysis technique, including the branching fraction fit,
by studying simulated $D\bar D$ events in a  sample 50 times the size 
of our data sample.
We reproduced the input branching fractions with biases due to our procedures
that were less than one-third of the statistical errors on the data and
consistent with zero.

\begin{table}[tb]
\caption{Fitted branching fractions and $D\bar D$ pair yields, along with the
fractional FSR corrections and comparisons to the Particle Data
Group~\protect{\cite{PDG}} fit results.  Uncertainties are
statistical and systematic, respectively.}
\label{tab-dataResults}
\begin{center}
\begin{tabular}{lccc}
\hline\hline
$D$ Decay Mode & Fitted ${\cal B}$ (\%) & PDG ${\cal B}$ (\%)
	& ~$\Delta_{\rm FSR}$\\
\hline
$\kpi$        & $3.91\pm 0.08\pm 0.09$        & $3.80\pm 0.09$ & ~$-2.0\%$ \\
$\kpipiz$     & $14.9\pm 0.3\pm 0.5$          & $13.0\pm 0.8$ & ~$-0.8\%$ \\
$\kpipipi$    & $8.3\pm 0.2\pm 0.3$           & $7.46\pm 0.31$ & ~$-1.7\%$ \\
\hline
$\kpipi$      & $9.5\pm 0.2\pm 0.3$           & $9.2\pm 0.6$ & ~$-2.2\%$ \\
$\kpipipiz$   & $6.0\pm 0.2\pm 0.2$           & $6.5\pm 1.1$ & ~$-0.6\%$ \\
$\kspi$       & $1.55\pm 0.05\pm 0.06$        & $1.41\pm 0.10$ & ~$-1.8\%$ \\
$\kspipiz$    & $7.2\pm 0.2\pm 0.4$           & $4.9\pm 1.5$ & ~$-0.8\%$ \\
$\kspipipi$   & $3.2\pm 0.1\pm 0.2$           & $3.6\pm 0.5$ & ~$-1.4\%$ \\
$\kkpi$       & $0.97\pm 0.04\pm 0.04$        & $0.89\pm 0.08$ & ~$-0.9\%$ \\
\hline\hline
$D\bar D$ Yield & \multicolumn{2}{c}{Fitted Value} & ~$\Delta_{\rm FSR}$ \\
\hline
$\NDzDzbar$ & \multicolumn{2}{c}{$(2.01\pm 0.04\pm 0.02)\times 10^5$} &
	~$-0.2\%$ \\
$\NDpDm$ & \multicolumn{2}{c}{$(1.56\pm 0.04\pm 0.01)\times 10^5$} &
	~$-0.2\%$ \\
\hline\hline
\end{tabular}
\end{center}
\end{table}

\begin{table}[tb]
\caption{Fitted ratios of branching fractions to the reference
branching fractions ${\cal R}_0\equiv {\cal B}(\Dzkpi)$ and
${\cal R}_+\equiv {\cal B}(\Dpkpipi)$, along with the
fractional FSR corrections and comparisions to the Particle Data
Group~\protect{\cite{PDG}} fit results.
Uncertainties are statistical and systematic, respectively.}
\label{tab-dataResultsRatios}
\begin{center}
\begin{tabular}{lccc}
\hline\hline
$D$ Decay Mode & Fitted ${\cal B}/{\cal R}_{0/+}$ & PDG ${\cal B}/{\cal R}_{0/+}$ & $\Delta_{\rm FSR}$\\
\hline
$\kpipiz$ & $3.82\pm 0.05\pm 0.10$ & $3.42\pm 0.22$ & $+1.2\%$ \\
$\kpipipi$ & $2.12\pm 0.03\pm 0.06$ & $1.96\pm 0.08$ & $+0.3\%$ \\
\hline
$\kpipipiz$  & $0.634\pm 0.014\pm 0.018$ & $0.70\pm 0.12$ & $+1.7\%$\\
$\kspi$ & $0.162\pm 0.004\pm 0.006$ & $0.153\pm 0.003$    & $+0.4\%$ \\
$\kspipiz$ & $0.753\pm 0.016\pm 0.039$ & --- & $+1.4\%$ \\
$\kspipipi$ & $0.336\pm 0.009\pm 0.014$ & $0.39\pm 0.05$  & $+0.8\%$ \\
$\kkpi$ & $0.102\pm 0.004\pm 0.003$ & $0.097\pm 0.006$    & $+1.3\%$ \\
\hline\hline
\end{tabular}
\end{center}
\end{table}

The results of the data fit are shown in Table~\ref{tab-dataResults}.
The $\chi^2$ of the fit is
28.1 for 52 degrees of freedom, corresponding to a confidence level of 99.7\%.
To obtain the separate contributions from statistical and systematic
uncertainties, we repeat the fit without any systematic inputs and take the
quadrature difference of uncertainties.
All nine branching fractions are consistent with, and most are higher than,
the current PDG averages~\cite{PDG}.  
In the $D$ candidate reconstruction, we do not explicitly search for FSR
photons. However, because FSR is simulated in the samples used to calculate
efficiencies, our measurements represent inclusive branching fractions for
signal processes with any number of photons radiated from the final state
particles.  If no FSR were included in the simulations, then all the branching
fractions would change by $\Delta_{\rm FSR}$ in Table~\ref{tab-dataResults}.

The correlation coefficient, $\rho$, between $\NDzDzbar$ and $\NDpDm$ is 0.07,
and each is essentially uncorrelated with branching fractions of the other
charge.  
Correlations among branching fractions are in the range 0.2--0.7.
In the absence of
systematic uncertainties, there would be almost no correlation between the
charged and neutral $D$ parameters.

We also compute ratios of branching fractions to the reference
branching fractions, shown in Table~\ref{tab-dataResultsRatios}.  These
ratios have higher precision than the individual
branching fractions, and they also agree with the PDG averages.  Without FSR
corrections to the efficiencies, all seven ratios would be 0.3\% to 1.7\%
higher.

We obtain the $e^+e^-\to D\bar D$ cross sections by scaling
$\NDzDzbar$ and $\NDpDm$ by the luminosity, which we determine to be
${\cal L} = (55.8 \pm 0.6)$ ${\rm pb}^{-1}$.  Thus, at
$E_{\rm cm}=3773$ MeV, we find peak cross sections of
$\sigma( e^+e^-\to D^0\bar D^0 ) = (3.60\pm 0.07^{+0.07}_{-0.05}) \ {\rm nb}$,
$\sigma( e^+e^-\to D^+ D^- ) = (2.79\pm 0.07^{+0.10}_{-0.04}) \ {\rm nb}$,
$\sigma( e^+e^-\to D\bar D ) = (6.39\pm 0.10^{+0.17}_{-0.08}) \ {\rm nb}$, and
$\sigma( e^+e^-\to D^+ D^- ) / \sigma( e^+e^-\to D^0\bar D^0 ) = 0.776\pm 0.024^{+0.014}_{-0.006}$,
where the uncertainties are statistical and systematic, respectively.
In addition to the systematic uncertainties on $\NDzDzbar$, $\NDpDm$, and the
luminosity, the cross section systematic uncertainties also include the effect
of $E_{\rm cm}$ variations with respect to the peak.
We account for the correlation between the charged and neutral cross sections
in computing the uncertainty on the total cross section.
Our measured
cross sections are in good agreement with BES~\cite{bessigma} and higher than
those of MARK III~\cite{markiii-2}.

In summary, we report measurements of three $D^0$ and six $D^+$ branching
fractions and the production cross sections $\sigma(D^0\bar D^0)$,
$\sigma(D^+D^-)$, and $\sigma(D\bar D)$
using a sample of 55.8 ${\rm pb}^{-1}$ of $e^+e^-\to D\bar D$ data
obtained at $E_{\rm cm} = 3773$~MeV.
We find branching fractions in agreement with, but somewhat higher, than
those in the PDG~\cite{PDG} summary.
We note that, unlike the branching fractions used
in the PDG averages, our measurements are corrected for FSR.
Not doing so would lower our branching fractions by 0.6\% to 2.2\%.
With our current data sample, the statistical and systematic uncertainties
are of comparable size.
Many of the systematic uncertainties, such as those for tracking 
and particle identification efficiencies, will be improved with larger
data samples.

We gratefully acknowledge the effort of the CESR staff 
in providing us with excellent luminosity and running conditions.
This work was supported by the National Science Foundation
and the U.S. Department of Energy.

\end{document}